\documentclass[twocolumn]{aastex63}
\usepackage[utf8]{inputenc} 
\usepackage[T1]{fontenc}    
\usepackage{hyperref}       
\usepackage{url}            
\usepackage{booktabs}       
\usepackage{amsfonts}       
\usepackage{nicefrac}       
\usepackage{microtype}      
\usepackage{amsmath}
\usepackage{ragged2e}
\usepackage{float}
\usepackage{graphicx}
\usepackage{natbib}
\usepackage[english]{babel}
\usepackage[normalem]{ulem}
\usepackage{cancel}
\usepackage{lineno}

\renewcommand{\vec}[1]{\boldsymbol{#1}}

\graphicspath{{./}{figures/}}
\begin{document}

\title{The Stability of the Electron Strahl against the Oblique Fast-magnetosonic/Whistler Instability in the Inner Heliosphere}

\author{Seong-Yeop Jeong}
\affiliation{Mullard Space Science Laboratory, University College London, Dorking, RH5 6NT, UK; s.jeong.17@ucl.ac.uk}

\author{Joel B. Abraham}
\affiliation{Mullard Space Science Laboratory, University College London, Dorking, RH5 6NT, UK; s.jeong.17@ucl.ac.uk}

\author{Daniel Verscharen}
\affiliation{Mullard Space Science Laboratory, University College London, Dorking, RH5 6NT, UK; s.jeong.17@ucl.ac.uk}

\author{Laura Ber\v{c}i\v{c}}
\affiliation{Mullard Space Science Laboratory, University College London, Dorking, RH5 6NT, UK; s.jeong.17@ucl.ac.uk}

\author{David Stansby}\affiliation{Mullard Space Science Laboratory, University College London, Dorking, RH5 6NT, UK; s.jeong.17@ucl.ac.uk}

\author{Georgios Nicolaou}
\affiliation{Mullard Space Science Laboratory, University College London, Dorking, RH5 6NT, UK; s.jeong.17@ucl.ac.uk}

\author{Christopher J. Owen}
\affiliation{Mullard Space Science Laboratory, University College London, Dorking, RH5 6NT, UK; s.jeong.17@ucl.ac.uk}

\author{Robert T. Wicks}
\affiliation{Northumbria University, Newcastle, NE1 8ST, UK}

\author{Andrew N. Fazakerley}
\affiliation{Mullard Space Science Laboratory, University College London, Dorking, RH5 6NT, UK; s.jeong.17@ucl.ac.uk}

\author{Jeffersson A. Agudelo Rueda}
\affiliation{Mullard Space Science Laboratory, University College London, Dorking, RH5 6NT, UK; s.jeong.17@ucl.ac.uk}

\author{Mayur Bakrania}
\affiliation{Mullard Space Science Laboratory, University College London, Dorking, RH5 6NT, UK; s.jeong.17@ucl.ac.uk}


\begin{abstract}
We analyze the micro-kinetic stability of the electron strahl in the solar wind depending on heliocentric distance. The oblique fast-magnetosonic/whistler (FM/W) instability has emerged in the literature as a key candidate mechanism for the effective scattering of the electron strahl into the electron halo population. Using data from Parker Solar Probe (PSP) and Helios, we compare the measured strahl properties with the analytical thresholds for the oblique FM/W instability in the low- and high-$\beta_{\parallel c}$ regimes, where $\beta_{\parallel c}$ is the ratio of the core parallel thermal pressure to the magnetic pressure. Our PSP and Helios data show that the electron strahl is on average stable against the oblique FM/W instability in the inner heliosphere. Our analysis suggests that the instability, if at all, can only be excited sporadically and on short timescales. We discuss the caveats of our analysis and potential alternative explanations for the observed scattering of the electron strahl in the solar wind. Furthermore, we recommend the numerical evaluation of the stability of individual distributions in the future to account for any uncertainties in the validity of the analytical expressions for the instability thresholds.\\
\\
\textit{Unified Astronomy Thesaurus concepts}: Solar wind (1534); Space plasma (1544); Heliosphere (711);
\end{abstract}



\section{Introduction} \label{intro}
The solar-wind plasma is continuously emitted from the Sun into interplanetary space \citep{1958ApJ...128..664P,1962Sci...138.1095N}. It expands quasi-spherically and forms the heliosphere. The electrons are the most abundant particle species and play an important role in the solar wind. They guarantee quasi-neutrality and provide the solar wind with a significant heat flux through the non-thermal properties of the electron velocity distribution functions \citep[VDFs;][]{2006LRSP....3....1M}. In addition, the electrons generate a global ambipolar electric field through their thermal pressure gradient \citep{1970A&A.....6..219J,LEMAIRE1970103,https://doi.org/10.1029/JA076i031p07479,https://doi.org/10.1029/1999JA900169,2019,2021}.  

According to previous \textit{in-situ} measurements of the electron VDF in the solar wind, multiple deviations from a Maxwellian equilibrium state often occur. The electron VDFs typically consist of three different electron populations: the core, the halo, and the strahl. The core, which accounts for over 90\% of the electrons in the solar wind, has a relatively low energy ($\lesssim 50\,\mathrm{eV}$) compared to the other populations and is quasi-isotropic. The  halo has a greater energy ($\gtrsim50\,\mathrm{eV}$) than the  core and is nearly isotropic as well \citep{https://doi.org/10.1029/2007JA012733,salem2021precision}. Lastly, the electron VDF often shows  an energetic and highly field-aligned  beam population known as the strahl. The  strahl carries most of the heat flux \citep{https://doi.org/10.1029/JA080i031p04181,https://doi.org/10.1029/JA092iA02p01103} and becomes faint at large  heliocentric distances \citep{doi:10.1029/JA092iA02p01075,doi:10.1029/2008JA013883,doi:10.1002/2016JA023656}. The continuing transfer of electrons from the strahl population into the halo population with increasing distance suggests that a local scattering mechanism modifies the shape of the VDF. This mechanism is very important for the definition of the electron heat flux; however, its nature is still unknown.

Electron measurements in the solar wind provide  evidence that  whistler waves exchange energy with the electron strahl, which makes them a candidate to explain the scattering of the  strahl into the halo population \citep{doi:10.1029/2006JA011967,2014ApJ...796....5L,doi:10.1002/2016JA023656}. Linear theory suggests that the FM/W wave propagating in the anti-Sunward direction with an  oblique wavevector with respect to the background magnetic field can be driven unstable by the strahl through the anomalous cyclotron resonance and scatter  strahl electrons in pitch angle through velocity space \citep{2019ApJ...871L..29V,2019ApJ...886..136V}. This strahl-driven instability has recently received a significant amount of attention in the literature as a local strahl scattering mechanism. Linear and quasi-linear theories as well as numerical  particle-in-cell (PIC) simulations support this picture \citep{L_pez_2020,Jeong_2020,Micera_2020,micera2021role,Sun_2021}.

As the solar wind begins its journey in the  corona where Coulomb collisions are more frequent than in the heliosphere, we anticipate that the core--halo--strahl configuration develops outside the corona. It is unknown at which heliocentric distance the local strahl scattering  sets in. Based on a  linear stability analysis, \citet{2018MNRAS.480.1499H} and \citet{2021MNRAS.507.1329S} argue against the action of a strahl-driven instability in the inner heliosphere at all. Parker Solar Probe (PSP) data from encounters 4 and 5 suggest that the  strahl, at times, drives the oblique FM/W wave unstable at heliocentric distances below $54r_s$ \citep{refId00}, where $r_s$ is the solar radius. A kinetic expansion model for the solar-wind electrons suggests, however, that the oblique FM/W instability is unlikely to occur regularly and scatter the electron strahl at heliocentric distances below $20r_s$ \citep{jeong2022kinetic}. Moreover, whistler waves are barely detected by PSP within $28r_s$ \citep{Cattell_2022}, supporting the notion that strahl scattering by the oblique FM/W instability is not a universal -- or even common -- process in the inner heliosphere. We note, however, that the required whistler-wave amplitude for effective strahl scattering is small \citep{Jeong_2020,doi:10.1063/5.0003401} and thus could be hidden in the fluctuation spectrum of the background turbulence in the solar wind \citep{https://doi.org/10.1029/96JA01275,2013PhRvL.110v5002C,Tong_2019}.

In the present paper, we use the analytical thresholds for the oblique FM/W instability derived by \citet{2019ApJ...886..136V} to investigate the stability of the electron strahl in the inner heliosphere. By using PSP and Helios data, we quantify the analytical thresholds as functions of heliocentric distance. Our results show that, on average, the electron strahl is not scattered by the oblique  FM/W instability in the inner heliosphere.


\section{Methods} \label{2}
\subsection{Thresholds for the Oblique FM/W Instability}\label{2.1}
To analyze the strahl scattering through the oblique FM/W instability, we evaluate the analytical thresholds in the low- and high-$\beta_{\parallel c}$ regimes  provided by \citet{2019ApJ...886..136V}. These analytical equations have been tested against numerical solutions for the thresholds of the oblique FM/W instability from linear hot-plasma theory over a range of representative solar-wind parameters. The transition between the low-$\beta_{\parallel c}$ and high-$\beta_{\parallel c}$ regimes occurs at
\begin{equation}\label{betasep}
\beta_{\parallel c}\approx 0.25,    
\end{equation}
where $\beta_{\parallel j} \equiv 8\pi n_j k_B T_{\parallel j}/B_0^2$, $n_j$ is the density of population $j$, $T_{\parallel j}$ is the temperature of population $j$ parallel to the background magnetic field, $\vec B_0$ is the background magnetic field, and $k_B$ is the Boltzmann constant. The subscripts $e$, $c$ and $s$ indicate quantities related to the total electrons, the core and the strahl, respectively. The plasma crosses the analytical threshold of the oblique FM/W instability in the low-$\beta_{\parallel c}$ regime if
\begin{equation}\label{lowthres}
    U_s\gtrsim 3v_{\parallel th,c},
\end{equation}
where $v_{\parallel th,c}=\sqrt{2k_BT_{\parallel c}/m_e}$ is the core parallel thermal speed, $U_s$ is the strahl bulk speed, and $m_e$ is the mass of an electron. The plasma crosses the analytical threshold of the oblique FM/W instability in the high-$\beta_{\parallel c}$ regime if
\begin{equation}\label{highthres}
    U_s\gtrsim  \left[2\frac{n_s}{n_c}\sqrt{\frac{T_{\parallel s}}{T_{\parallel c}}} v_{Ae}^2 v_{\parallel th,c}^2 \frac{(1+\cos\theta)}{(1-\cos \theta)\cos \theta} \right]^{1/4},
\end{equation}
where  $v_{Ae}=B_0/\sqrt{4 \pi n_e m_e}$, and $\theta$ is the angle between wavevector $\vec k$ and  $\vec B_0$\footnote{The original derivation of the instability thresholds by \citet{2019ApJ...886..136V} assumes isotropic electrons. However, it is straight-forward to extend this framework to anisotropic electrons, in which case we retrieve  Eqs.~(\ref{lowthres}) and (\ref{highthres}).}.

\subsection{Data Analysis of PSP and Helios Data}\label{2.2}

We base our data analysis on the PSP dataset created by \cite{Abraham2021_submitted}. Our PSP data were recorded at heliocentric distances between $25r_s$ and $90r_s$. This dataset uses fits to the observed level-3 electron VDFs provided by the Solar Wind Electron Alphas and Protons (SWEAP) instrument suite \citep{2016SSRv..204..131K,2020ApJS..246...74W} on board PSP.  SWEAP consists of two electrostatic analysers, SPAN A and B,  which combined measure electrons arriving from across almost the full sky using orthogonally positioned $120^{\circ} \times 240^{\circ}$ fields of view, over an energy range from 2~eV to 1793~eV during our measurement intervals. During the encounter, the sampling cadence of PSP's SPAN detector is $13.98s$, and its integration time is $6.965s$. We fit the VDFs in the magnetic-field-aligned velocity frame and ignore any data points below 30~eV to avoid the inclusion of secondary electrons. The analysis method fits a bi-Maxwellian distribution to the core, a bi-$\kappa$-distribution to the halo and a drifting bi-Maxwellian distribution to the strahl. We only allow the strahl to drift in the parallel direction to the magnetic field.

On Helios 1 and 2 \citep{Rosenbauer1981,Pilipp1987}, electrons were measured by the electron particle instrument I2, part of the E1 Plasma Experiment. Our Helios data were recorded at heliocentric distances between $62r_s$ and $210r_s$. A narrow instrument aperture ($19^\circ \times 2^\circ$) spins with the spacecraft and covers a full range of $360^\circ$ in azimuth, providing two-dimensional electron VDFs. Helios took $16\,\mathrm s$ to measure a full two-dimensional electron VDF with a cadence of $40\,\mathrm s$. For context, we also use proton density and velocity as well as magnetic-field measurements as described by \citet[][Section~3]{Bercic2019}. We analyze the electron VDFs in the magnetic-field-aligned frame, in which the solar-wind protons have zero bulk speed. We model the distributions with a sum of a bi-Maxwellian distribution representing the core, a bi-$\kappa$-distribution representing the halo, and a bi-Maxwellian distribution representing the strahl. We only allow the core and the strahl to have drift velocities in the direction parallel to the magnetic field. 

In both datasets, we discard all VDFs for which the fit results for the density of the halo or the strahl is greater than the density of the core.

\section{Results} \label{3}

\subsection{Fit Parameters from PSP and Helios }\label{3.1}

\begin{figure}[ht]
  \centering
  \includegraphics[width=\columnwidth]{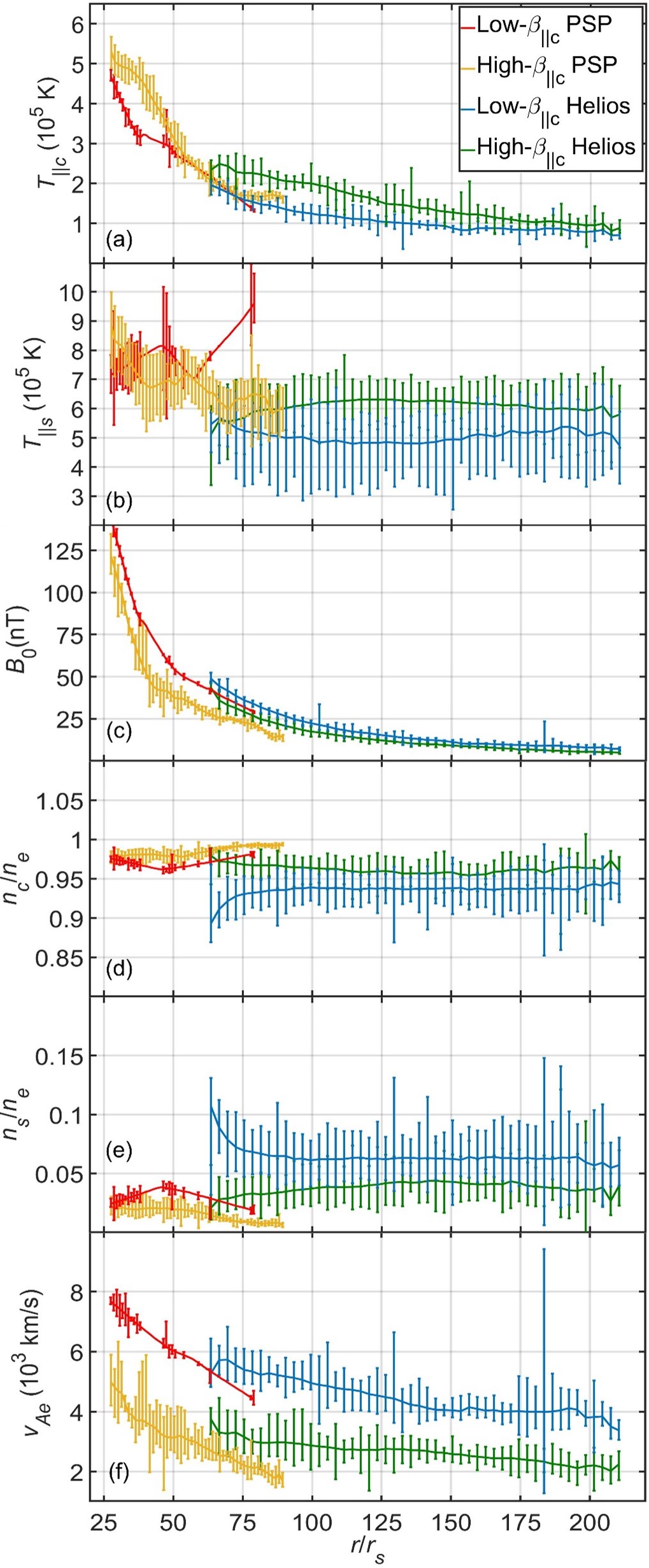}
  \caption{Radial profiles of $T_{\parallel c}$, $T_{\parallel s}$, $B_0$, $n_c/n_e$, $n_s/n_e$ and $v_{Ae}$ in our PSP and Helios datasets, separated between the low-$\beta_{\parallel c}$ and the high-$\beta_{\parallel c}$ regimes according to Eq.~(\ref{betasep}).}
  \label{profile}
\end{figure}

We bin the fitted data from both PSP and Helios  into 50 radial-distance bins of equal width. We first separate our PSP and Helios data into the low-$\beta_{\parallel c}$ regime and high-$\beta_{\parallel c}$ regime according to Eq.~(\ref{betasep}). We show the profiles of $T_{\parallel c}$, $T_{\parallel s}$, $B_0$, $n_c/n_e$, $n_s/n_e$ and $v_{Ae}$, which all appear on the right-hand sides of Eqs.~(\ref{lowthres}) and (\ref{highthres}), as functions of heliocentric distance in Fig.~\ref{profile}.
The red and blue lines correspond to the parameters in the low-$\beta_{\parallel c}$ regime while the yellow and green lines correspond to the parameters in the high-$\beta_{\parallel c}$ regime from PSP and Helios. 

PSP predominately samples streams of slow solar wind while the Helios dataset includes a broader variety of solar wind. Therefore, the PSP and Helios datasets do not connect exactly, especially in terms of $T_{\parallel s}$, $n_c/n_e$ and $n_s/n_e$ in the low-$\beta_{\parallel c}$ cases. Fig.~\ref{profile}(c) shows the magnetic field averaged over the respective particle sampling times of the electron instruments for PSP and Helios.

For our evaluation of the instability thresholds as functions of heliocentric distance, we apply the binned mean profiles of the parameters shown in Fig.~\ref{profile} to Eqs.~(\ref{lowthres}) and (\ref{highthres}).

\subsection{Threshold in the Low-$\beta_{\parallel c}$ Regime} \label{3.2}

\begin{figure}[ht]
  \centering
  \includegraphics[width=\columnwidth]{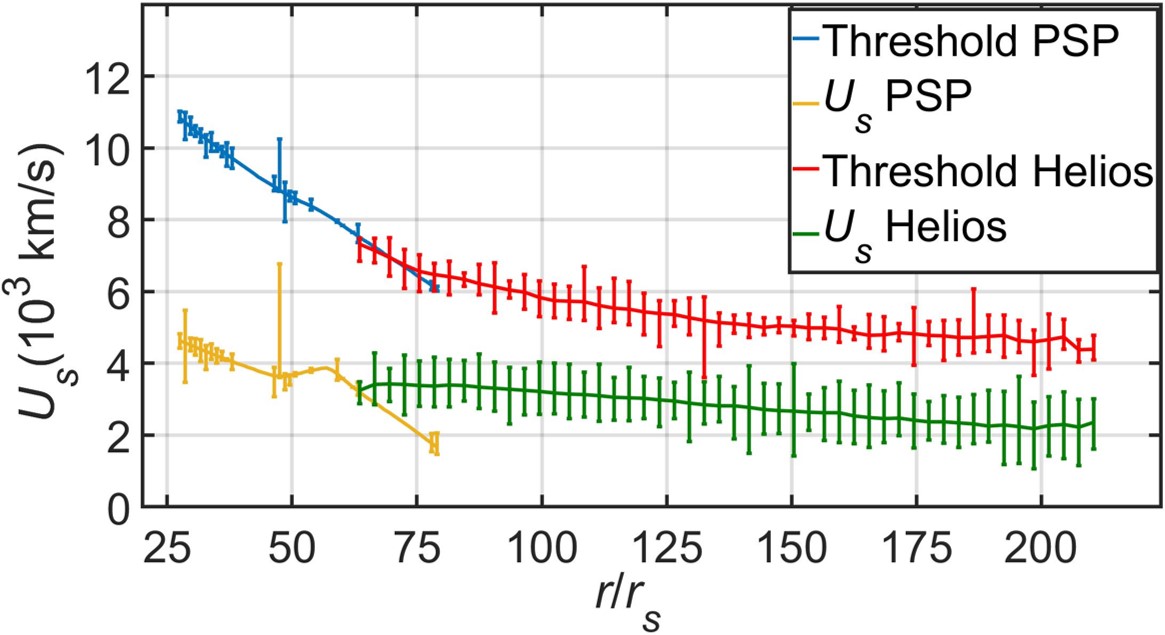}
  \caption{  Profiles of the low-$\beta_{\parallel c}$ threshold of the oblique FM/W instability from Eq.~(\ref{lowthres}) and $U_s$ as functions of heliocentric distance. All PSP and Helios data underlying this figure have $\beta_{\parallel c}<0.25$.}
  \label{lowbeta}
\end{figure}

In Fig.~\ref{lowbeta}, the blue (PSP) and red (Helios) lines show the profiles of the right-hand side of Eq.~(\ref{lowthres}), and the yellow (PSP) and green (Helios) lines show the measured profiles of $U_s$.
The strahl bulk speed does not cross the low-$\beta_{\parallel c}$ threshold in the shown range of heliocentric distances on average and, thus, does not fulfill Eq.~(\ref{lowthres}). The low-$\beta_{\parallel c}$ threshold and $U_s$ decrease quasi-monotonously. Near the Sun, the difference between the low-$\beta_{\parallel c}$ threshold and $U_s$ decreases with heliocentric distance. However, beyond  $\sim 80r_s$ from the Sun, the difference is nearly constant. The error bars of the low-$\beta_{\parallel c}$ threshold and $U_s$  do not overlap in the shown range of heliocentric distances.

\subsection{Threshold in the High-$\beta_{\parallel c}$ Regime} \label{3.3}

Previous studies of the oblique FM/W instability by \citet{2019ApJ...871L..29V}, \citet{2019ApJ...886..136V}, \citet{L_pez_2020}, \citet{Jeong_2020}, \citet{Micera_2020}, \citet{micera2021role}, and \citet{Sun_2021} suggest that the angle $\theta$ in Eq.~(\ref{highthres}) typically ranges from $\theta=55^\circ$ to $65^\circ$. Eq.~(\ref{highthres}) does not significantly depend on $\theta$ in this range compared to its much stronger dependence on $T_{\parallel c}$, $T_{\parallel s}$, $n_c$, $n_s$ and $v_{Ae}$. Therefore, we use a representative value of $\theta=60^\circ$.

\begin{figure}[ht]
  \centering
  \includegraphics[width=\columnwidth]{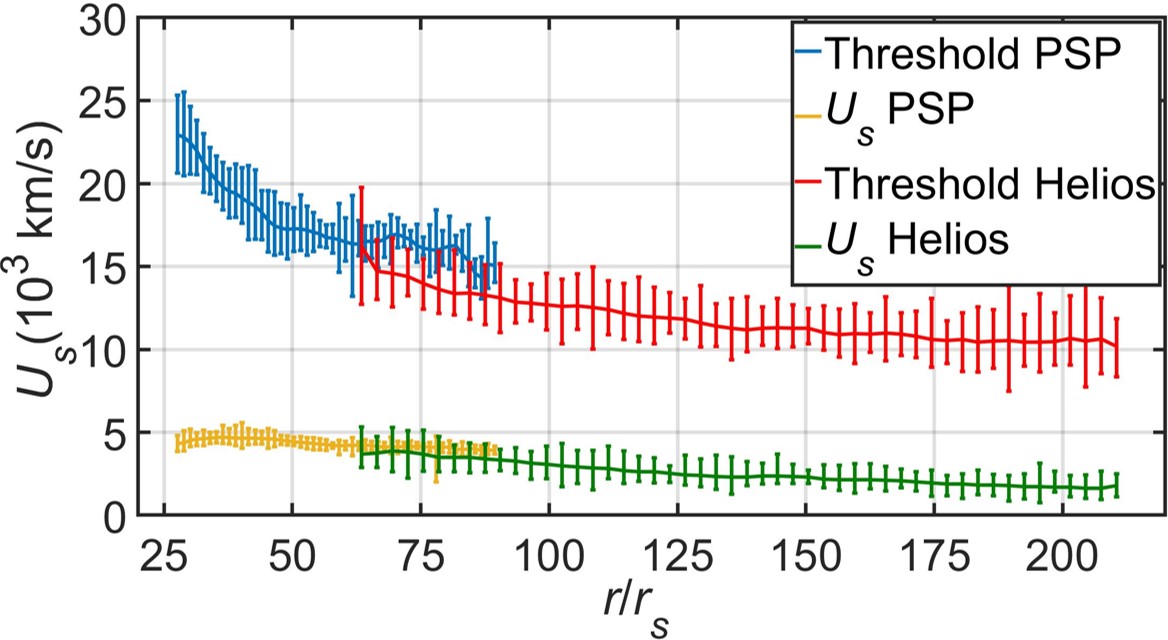}
  \caption{Profiles of the high-$\beta_{\parallel c}$ threshold of the oblique FM/W instability from Eq.~(\ref{highthres}) and $U_s$ as functions of heliocentric distance. All PSP and Helios data underlying this figure have $\beta_{\parallel c}>0.25$.}
  \label{highbeta}
\end{figure}

In Fig.~\ref{highbeta}, the blue (PSP) and red (Helios) lines show the profiles of the right-hand side of Eq.~(\ref{highthres}), and the yellow (PSP) and green (Helios) lines show the measured profile of $U_s$.
The strahl bulk speed does not cross the high-$\beta_{\parallel c}$ threshold in the shown range of heliocentric distances on average and, thus, does not fulfill Eq.~(\ref{highthres}). The high-$\beta_{\parallel c}$ threshold and $U_s$ quasi-monotonously decrease at all explored heliocentric distances. Like in the low-$\beta_{\parallel c}$ case, near the Sun, the difference between the high-$\beta_{\parallel c}$ threshold and $U_s$ decreases with heliocentric distance. However, beyond  $\sim 80r_s$ from the Sun, the difference is approximately constant. The error bars of the high-$\beta_{\parallel c}$ threshold and $U_s$ do not overlap  at all explored heliocentric distances. 

\section{Discussion and Conclusions} \label{4}

We analyze the stability of the solar-wind electron strahl against the oblique FM/W instability for self-induced strahl scattering in two different $\beta_{\parallel c}$ regimes. By quantifying the plasma parameters based on binned and averaged electron  data from PSP and Helios, as shown in Fig.~\ref{profile}, we evaluate the analytical instability thresholds as functions of  heliocentric distance. 

Our results show that Eqs.~(\ref{lowthres}) and (\ref{highthres}) are, for plasma with the observed average radial profiles, not satisfied  at heliocentric distances between  $25r_s$ and $210r_s$. In Figs.~\ref{lowbeta} and \ref{highbeta}, the low- and high-$\beta_{\parallel c}$ thresholds are always greater than $U_s$. For both the low-$\beta_{\parallel c}$ and for the high-$\beta_{\parallel c}$ cases, the difference between the thresholds and $U_s$ decreases with heliocentric distance near the Sun as predicted by \citet{jeong2022kinetic}. 
However, beyond a heliocentric distance of $\sim80r_s$, the difference between the thresholds and $U_s$ is nearly constant. Assuming that our PSP and Helios data are  representative  and that our Eqs.~(\ref{lowthres}) and (\ref{highthres}) reflect the thresholds accurately, our findings suggest that the  strahl is not scattered by the oblique FM/W instability in the inner heliosphere, in agreement with the arguments presented by \citet{2018MNRAS.480.1499H} and \citet{2021MNRAS.507.1329S}.

According to quasi-linear theory and particle-in-cell simulations \citep{Jeong_2020,Micera_2020,micera2021role}, the oblique FM/W instability, if excited, scatters the electron strahl efficiently on a timescale shorter than $1\,\mathrm s$ in the inner heliosphere. This rapid action of the instability and the radial decrease of the thresholds suggest that the measured $U_s$ would closely follow the threshold of marginal stability at all distances beyond the critical distance $r_{\mathrm{crit}}$, at which the threshold is crossed for the first time. Neither Fig.~\ref{lowbeta} nor Fig.~\ref{highbeta} suggest the existence of $r_{\mathrm{crit}}$ at heliocentric distances less than 1~au. However, we propose some uncertainties to our results.

Our analysis is based on electron measurements with a finite sampling cadence ($\sim 14\,\mathrm s$ by PSP and $\sim 40\,\mathrm s$ by Helios). Even if the instability scatters the electron strahl, PSP and Helios would be unlikely to detect the moment of strahl scattering because it occurs so quickly. If the magnetic field changes significantly during such a sampling interval, the corresponding electron measurement would be deteriorated. Since the averaged amplitude of magnetic-field variations at the timescale of the particle measurements is small, we assume that this effect has a small impact on our results. However, we recommend repeating our study with even faster plasma instrumentation \citep[e.g., with Solar Orbiter's EAS instrument in burst mode,][]{refId00000} for comparison in the future.

The derivation of Eqs.~(\ref{lowthres}) and (\ref{highthres}) by \citet{2019ApJ...886..136V} is based on the assumption that the oblique FM/W instability is most efficiently driven by electrons with  $v_\parallel=U_s$ where $v_\parallel$ is the velocity coordinate in the direction parallel with respect to $B_0$ in the solar-wind frame. However, according to the predictions from quasi-linear theory and  PIC simulations \citep{Jeong_2020,Micera_2020,micera2021role},  strahl scattering by the oblique FM/W instability most efficiently occurs in the $v_\parallel$-range $\gtrsim U_s$. For example, \citet{Jeong_2020} show that the oblique FM/W instability in the high-$\beta_{\parallel c}$ regime is most efficiently driven by electrons with  $v_\parallel \approx 2U_s$. Applying this argument in the derivation of Eqs.~(\ref{lowthres}) and (\ref{highthres}), we find a threshold for the high-$\beta_{\parallel c}$ case that is lower  by a factor of two than shown in Fig.~\ref{highbeta}. Such a modification would lead to a small difference between the two curves shown in Fig.~\ref{highbeta}.
If we adjusted the analytical thresholds by a constant factor greater than 2, the discovered constant difference between the thresholds and $U_s$ at $r\gtrsim 80r_s$  would be consistent with the limitation of $U_s$ through the oblique FM/W instability with $r_{\mathrm{crit}}\approx 80r_s$.

The uncertainties of the analytical instability thresholds can be circumvented by calculating the stability of individually measured electron VDFs with numerical tools like the ALPS \citep{Verscharen_2018} or LEOPARD  \citep{https://doi.org/10.1002/2016JA023522}. These numerical codes are able to calculate  the full hot-plasma dispersion relation based on our PSP and Helios data without assuming a bi-Maxwellian shape of the electron VDF. In addition, it would be worthwhile to investigate the occurrence of oblique FM/W waves as a result of the oblique FM/W instability. This method would also allow for an independent determination of $r_{\mathrm{crit}}$ and thus an independent determination of a potential correction factor to our analytical instability thresholds.

Our observation of the decreasing trends in $U_s$ (Figs.~\ref{lowbeta} and \ref{highbeta}) as well as earlier observations of the broadening of the strahl pitch-angle and the scattering of the strahl into the halo \citep{1996A&A...316..350H,https://doi.org/10.1029/2005JA011119,doi:10.1029/2008JA013883,2017JGRA..122.3858G} provide evidence that strahl scattering occurs over a wide range of distances in the solar wind. Therefore, if the oblique FM/W instability is \emph{not} responsible for this evolution, other processes must explain the observed trends.

Other linear instabilities, including the quasi-parallel whistler heat-flux instability \citep{Tong_2019,Kuzichev_2019,doi:10.1063/5.0003401} are not efficient strahl-scattering mechanisms \citep{2019ApJ...886..136V}. However, nonlinear processes involving combinations of primary and secondary instabilities in connection with solar-wind expansion reproduce observed wave and particle features in the solar wind \citep{micera2021role}. Coulomb collisions have been suggested to  contribute to the strahl evolution even though the collisional mean free path of suprathermal electrons is large \citep{2012ApJ...760..143L,2018MNRAS.474..115H,2019MNRAS.484.2474H}. Another possible strahl-scattering mechanism involves the dissipation of existing plasma fluctuations \citep[especially whistler waves,][]{2012SSRv..172..303V,Vocks_2005}. Although whistler waves have been sporadically observed in the solar wind \citep{Lacombe_2014,Agapitov_2020,refId000,refId0,refId0000}, their origin and possible links to  solar-wind turbulence are still not well understood.

\acknowledgments

D.V.~is supported by STFC Ernest Rutherford Fellowship ST/P003826/1. J. B. A. is supported by STFC grant ST/T506485/1. D.V., C.J.O., A.N.F., D.S., and L.B.~are supported by STFC Consolidated Grant ST/S000240/1. R. T. W. is funded by STFC consolidated Grant ST/V006320/1. J.A.A.R.~is supported by the ESA Networking/Partnering Initiative (NPI) contract 4000127929/19/NL/MH/mg and the Colombian programme Pasaporte a la Ciencia, Foco Sociedad -- Reto 3, ICETEX grant 3933061. M. B. is supported by a UCL Impact Studentship, joint funded by the ESA NPI contract 4000125082/18/NL/MH/ic. This work was discussed at the ``Joint Electron Project'' at MSSL. The authors acknowledge insightful discussions within the International Team ``Heliospheric Energy Budget: From Kinetic Scales to Global Solar Wind Dynamics'' at the International Space Science Institute (ISSI) in Bern led by M.~E. Innocenti and A.~Tenerani.

\bibliography{sample63}{}
\bibliographystyle{aasjournal}



\end{document}